# GIVE QUANTUM MECHANICS A CHANCE: USE RELATIVISTIC QUANTUM MECHANICS TO ANALYZE MEASUREMENT!

Karl-Erik Eriksson  <frtkee@chalmers.se>
Complex Systems Group, Department of Space, Earth and Environment, Chalmers University of Technology, SE-41296 Gothenburg, Sweden

Abstract

At the time of publication of H. Everett's Relative-State Formulation (1957) and DeWitt's Many-Worlds Interpretation (1970), quantum mechanics was available in a more modern and adequate version than the one used by these authors. We show that with the more modern quantum theory, quantum measurement could have been analyzed along more conventional lines in a one-world cosmology. Bell criticized the Everett-DeWitt theory quite sharply in 1987 but this seems not to have affected the acceptance of the old quantum mechanics as the framework for analysis of measurement.

## 1. INTRODUCTION

Resignation with respect to the possibility of solving the measurement problem is now a very common attitude among physicists and it has been so for a long time. Already in his famous 'Feynman Lectures', Richard Feynman [1] expressed this kind of resignation, not without regret:

> [P]hysics has given up on the problem of trying to predict exactly what will happen in a definite circumstance. Yes! physics has given up. *We do not know how to predict what would happen in a given circumstance*, and we believe now that it is impossible, that the only thing that can be predicted is the probability of different events. It must be recognized that this is a retrenchment in our earlier ideal of understanding nature. It may be a backward step, but no one has seen a way to avoid it. [[1], Chapter 37, p. 37-10.]

It is somewhat ironic that one way, suggested by John Bell [2], to avoid the "backward step" (the quotation will come in Section 3.2), is a work of Feynman, his 'sum over all possible paths'. In our discussion, we do not use this theory but we use Feynman's closely related work in Quantum Field Theory. This changes completely the whole picture of interaction in measurement and opens a new way of analysis.

The theoretical methods of Quantum Field Theory were available already at the time of Hugh Everett's work [3] on his Relative-State Formulation and even more developed at the time of Bryce DeWitt's paper [4] on the Many-Worlds Interpretation.

In Section 2, we briefly review the well-known background: von Neumann's dual time development in quatum mechanics, Schrödinger's cat, Everett's Relative-State Formulation and DeWitt's Many-Worlds Interpretation. We then identify the 1930 quantum mechanics as an inadequate theoretical basis for all this.

In Section 3, we specify some criticism against basic assumptions in the reviewed background. We also quote Bell's criticism from 1987 [2] of Everett [3] and DeWitt [4], a criticism of the lack of reversibility in their theory. Bell's criticism seems to have been largely ignored in the discussion on measurement.

In Section 4, we quote Steven Weinberg and Brian Greene from their reviews of the measurement problem and their expressions of discontent. We also mention two of the unromantic approaches to quantum measurement in Bell's typology, the pragmatic approach and the non-linear/stochastic approach.

Bell's criticism of the Reative-State Formulation and the Many-Worlds Interpretation is a good reason to turn from the non-relativistic quantum mechanics of the 1930s to relativistic quantum mechanics, i.e., Quantum Field Theory. In Section 5, we describe this choice and present a generic model within the scattering theory of Quantum Field Theory. In this model, the interaction between the measured system and the part of the measuring apparatus that it first meets, looks totally different from that of the traditional non-relativistic theory. In our model, the included part of the apparatus is allowed to influence the scattering amplitudes. A simple statistical analysis of this influence reveals a bifurcation of the kind that we know from measurement with relative frequences of outcomes that agree with the Born rule.

In the concluding section, we quote Bell on the large general interest in the "romantic" approaches to measurement as compared to unromantic approaches preferred by him and by us. The difficulty to see the connection between Quantum Field Theory and measurement has been staying with the physics community for a very long time.

## 2. BACKGROUND

### 2.1. von Neumann's dilemma and Schrödinger's cat

John von Neumann could not avoid having two different time developments in quantum theory. Andrew Whitaker [5] has described the situation as follows:

> For von Neumann [...], the collapse of the wave function seemed little more than a statement of what a measurement is. Yet he admitted that it caused great conceptual problems. In his words, there appears to be 'a peculiar dual role of the quantum mechanical procedure'. In the absence of any measurement, the wavefunction develops according to the Schrödinger equation (a process of type 2, as von Neumann called it); at a measurement, it follows the projection postulate (a process of type 1).

There are very considerable mathematical differences between processes of type 1 and type 2. [...] [I]t is easier to describe the differences physically, thermodynamically in fact. Von Neumann himself showed that a process of type 2 is *thermodynamically reversible*; it remains possible at the end of the process to restore the system to its initial state. For a process of type 1, no such possibility exists, and the process is called *thermodynamically irreversible*.

Despite this apparent dichotomy, there have been innumerable attempts to demonstrate that a process of type 1 could be approximately, or in some limit, equivalent to one of type 2. [...]

Indeed, the whole area has become known as the *measurement problem* of quantum theory. It is especially embarrasing because, as Bell in particular stressed, one should really be able to describe the 'measurement' in terms of straightforward quantum-mechanical processes of type 2 of the atoms of the measuring device. So how can the measurement itself be of type 1? [Whitaker [5], p. 196]

To describe the problem, we take the measured system $\mu$ to be a two-level system such as the spin of an electron and consider measurement of the vertical spin direction $\sigma_z$. A measurement of $\sigma_z$ on $\mu$ with $\mu$ remaining after the process (a non-destructive measurement), gives as result one of the eigenvalues +1 or −1, and the process takes $\mu$ into the corresponding eigenstate: when the measurement result is +1, $\mu$ is taken into the state $|+\rangle_\mu$ and when the measurement result is −1, $\mu$ is taken into the state $|-\rangle_\mu$.

For a single measurement, the result can not be predicted unless $\mu$ is already in an eigenstate of $\sigma_z$. However, statistically, we know that for a given initial state of $\mu$,

$$|\psi\rangle_\mu = \psi_+ |+\rangle_\mu + \psi_- |-\rangle_\mu \quad (|\psi_+|^2 + |\psi_-|^2 = 1), \tag{2.1}$$

the + result obtains with the probability $|\psi_+|^2$ and the − result obtains with the probability $|\psi_-|^2$ (the Born rule).

We shall consider the interaction between $\mu$ and a very limited part $A$ of the measurement apparatus. In early attempts to analyze measurement, one assumed $A$ to be in some given initial state $|i\rangle_A$. Knowing that the measurement process would not change initial $\sigma_z$ eigenstates of $\mu$, $|+\rangle_\mu$ or $|-\rangle_\mu$, one concluded that a measurement process would take the initial state

$$|+\rangle_\mu \otimes |i\rangle_A \quad \text{or} \quad |-\rangle_\mu \otimes |i\rangle_A, \tag{2.2}$$

of $\mu \cup A$ into the final state,

$$|+\rangle_\mu \otimes |f,+\rangle_A \quad \text{or} \quad |-\rangle_\mu \otimes |f,-\rangle_A, \tag{2.3}$$

respectively, with $|f,+\rangle_A$ and $|f,-\rangle_A$ labelling final states of $A$, beginning to register the result + or −.

Then because of the linearity of quantum dynamics, the same kind of process would carry the initial superposition state

$$|\psi\rangle_\mu \otimes |i\rangle_A = \psi_+ |+\rangle_\mu \otimes |i\rangle_A + \psi_- |-\rangle_\mu \otimes |i\rangle_A \tag{2.4}$$

into the final state

$$\psi_+ |+\rangle_\mu \otimes |f,+\rangle_A + \psi_- |-\rangle_\mu \otimes |f,-\rangle_A. \tag{2.5}$$

Here one has achieved entanglement between $\mu$ and $A$, but no step is taken towards a definite measurement result. This problem dates back to John von Neumann and, following Michael Nauenberg [6], we shall call it <u>von Neumann's dilemma</u>.

We note that in (2.2) the same initial state $|i\rangle_A$ for $A$ could lead to both final results in (2.3). But this does not have to be the case; the transition rates to the final states could depend crucially on the initial state of $A$ and lead us to a situation that is different from (2.5) with von Neumann's dilemma. However, in the dynamical description used, there are not even any transition rates. This points towards the necessity of going to a more modern version of quantum mechanics: <u>relativistic quantum mechanics</u>.

Erwin Schrödinger's Gedankenexperiment [7] with a cat continued this kind of discussion. Through the measurement kind of interaction, the small quantum system was allowed to entangle itself with a much larger system, including the cat, and drag it into the superposition without changing the coefficients. This brings the cat into the famous superposition of two states, one with a living cat and the other with a dead cat. Closed-system quantum development was considered to be applicable without any restriction.

2.2. <u>The Relative-State Formulation and The Many-Worlds Interpretation</u>

To get away from the projection-type development for measurement (von Neumann's type-1 process), Hugh Everett [3] suggested The Relative-State Formulation (RSF) in 1957 with only type-2 processes. As we have seen in the discussion above, this leads to entanglement but not to 'reduction', i.e.

all outcomes are open. However, since there is entanglement, in each term of the superposition there is agreement between the observed system and what is recorded by the system $A$.

Bryce DeWitt [4] transformed Everett's formulation into a complete world view, the Many-Worlds Interpretation (MWI) in 1970. He presented this world view in close connection to a personal description of his own reaction to it:

[E]very quantum transition taking place on every star, in every galaxy, in every remote corner of the universe is splitting our local world on earth into myriads of copies of itself.

His hesitation was immediate:

I still recall vividly the shock I experienced on first encountering this multiworld concept. The idea of $10^{100+}$ slightly imperfect copies of oneself all constantly splitting into further copies, which ultimately become unrecognizable, is not easy to reconcile with common sense.

The point taken here is that the more developed relativistic quantum theory opens possibilities that are easier to reconcile with common sense and with a single world.

2.3. Basic theory behind RSF and MWI: non-relativistic quantum mechanics

The non-relativistic (and non-reversible) quantum dynamics of the early 1930s was insufficient for a deeper analysis of measurement and it led Everett and DeWitt to a theory without a definite measurement result. What seems more surprising is that still to this day, the same version of quantum dynamics continues to be accepted as the framework for discussion about the foundations of quantum mechanics.

Already at the time of Everett's and DeWitt's work, the more modern relativistic quantum mechanics had been developed and very successfully used, for instance in Quantum Electrodynamics. As we shall see, the newer quantum mechanics could have been used to make a totally different and more correct analysis of measurement.

3. CRITICISM OF THE MENTIONED THEORIES

3.1. Criticism of von Neumann's dilemma and Schrödinger's cat

In RSF and MWI, it is implicitly assumed that the same initial state of the apparatus can lead to any of the measurement results. We have considered interaction between the measured system $\mu$ with a (limited) part of the apparatus $A$. It is clear that $A$ must have many degrees of freedom and hence a very large ensemble of possible initial states. As Everett and DeWitt have done, we can consider $\mu \cup A$ to evolve as a closed quantum-mechanical system. However, the discussion here will be a one-world discussion. Then with $\mu$ in a given initial state

$$|\psi\rangle_\mu = \psi_+|+\rangle_\mu + \psi_-|-\rangle_\mu, \tag{3.1}$$

different outcomes from its interaction with $A$, <u>must</u> depend on different initial states of $A$. The strength of the measurement interactions must depend on the initial state of $A$; we should be open for the possibility that this dependence can be quite strong. The situation in each single measurement process is unique. In the mean, however, $A$ must be unbiased with respect to the outcomes. In the old theory, there was no room for describing interaction strength.

In his Gedankenexperiment, Schrödinger [7] extended the limit for what can be a <u>closed</u> system to be described by internal deterministic quantum mechanics. In common scientific thinking, living systems, such as cats, are by their very nature, open systems, but the physics culture seems to allow Schrödinger's cat to remain an eternal exception.

If one stays in <u>one world</u> and considers the system that includes the cat as an <u>open</u> system, the whole <u>Gedankenexperiment</u> breaks down and the discussion becomes pointless.

3.2. <u>Bell's criticism of RSF/MWI</u> (J.S. Bell 1987)

The development described in RSF/MWI was criticized by John Bell [2] for <u>not</u> having the usual characteristics of a quantum theory:

Thus DeWitt seems to share our idea that the fundamental concepts of the theory should be meaningful on a microscopic level and not only on some ill-defined macroscopic level. But at the microscopic level there is no such asymmetry in time as would be indicated by the existence of branching and the non-existence of debranching. [...] [I]t even seems reasonable to regard the coalescence of previously different branches, and the resulting interference phenomena, as *the* characteristic feature of quantum mechanics. In this respect an accurate picture, which does not have any tree-like character, is the 'sum over all possible paths' of Feynman. [[2], Chapter 15, p. 135]

Thus Bell's criticism is that the Everett-DeWitt theory misses an essential feature of quantum mechanics which we can call reversibility. By this we mean that if one kind of elementary process is taking part in an overall process, then the inverse elementary process should also be included. This is another sign that one has to go beyond the methods of the 1930s and proceed to the manifestly reversible relativistic quantum mechanics of the 1950s, i.e., Quantum Field Theory (QFT), as a basis for analyzing measurement. We consider Bell's criticism to be decisive: the quantum mechanics of the early 1930s is <u>not</u> adequate.

3.3. <u>Relativistic quantum mechanics</u>

From the late 1930s to the mid 1950s, relativistic quantum mechanics developed into QFT, sufficiently good for very accurate dynamical predictions in Quantum Electrodynamics, <u>and</u> sufficiently good for general quantum-mechanical considerations applied to particle physics. Going from non-relativistic to relativistic theory involves more than taking into account relativistic effects due to fast motion of particles. It changes the whole picture of particle physics. Non-relativistic and relativistic quantum theory are structurally different; the non-relativistic theory is much less than the limit $\frac{|p|}{mc} \ll 1$ of the relativistic theory.

In the 1950s, important comparisons between theory and experiment took place within particle physics and Quantum Electrodynmics. It is fair to say that the theory tested was relativistic quantum mechanics. In this sense, <u>quantum mechanics was relativistic quantum mechanics, i.e., Quantum Field Theory</u> (QFT). Since this time, relativistic quantum mechanics has continued to be the framework for the physics of elementary particles and their interactions.

Still to this day, within the physics community, discussions about the foundations of quantum mechanics, including measurement, usually take place within a pre-relativistic framework. I shall return to this issue.

Since the connection of the final states of $\mu \cup A$ to their initial states is of crucial interest for the measurement process, <u>scattering theory of QFT</u> should be a suitable theoretical framework for describing the process. Essential concepts in this theory are transition amplitudes and transition rates, sufficient for us when details of the $\mu A$-interaction have to remain unknown.

Relativistic quantum mechanics was certainly available in Princeton the years before Everett's paper. If it had been used as a basis for the analysis of interaction in a measurement, Hugh Everett and John Wheeler could have drawn conclusions totally different from those drawn in RSF or MWI.

In our discussion below, we shall only use knowledge that was available in the mid-1950s.

# 4. DISCONTENT WITH THE SITUATION IN QUANTUM MEASUREMENT

## 4.1. Weinberg's discontent

Not only Feynman has expressed discontent concerning the situation with the measurement problem. A few years ago, Steven Weinberg wrote a popular article in the New York Review of Books [8]. He started in a MWI description:

> [I]n consequence of their interaction during measurement, the wave function becomes a superposition of two terms, in one of which the electron spin is positive and everyone in the world who looks into it thinks it is positive, and in the other the spin is negative and everyone thinks it is negative. Since in each term of the wave function everyone shares the belief that the spin has one definite sign, the existence of the superposition is undetectable. In effect the history of the world has split into two streams, uncorrelated with each other.
>
> This is strange enough, but the fission of history would not only occur when someone measures a spin. In the realist approach the history of the world is endlessly splitting; it does so every time a macroscopic body becomes tied in with a choice of quantum states. This inconceivably huge variety of histories has provided material for science fiction, and it offers a rationale for a multiverse[.]

However, Weinberg was not satisfied with this situation; he would prefer a one-world theory:

> But the vista of all these parallel histories is deeply unsettling, and like many other physicists I would prefer a single history.

Research on quantum measurement is still going on but it is probably correct to say that resignation with respect to solving the measurement problem in a single world, is a dominant standpoint in the culture of the physics community.

## 4.2. Greene's discontent

A few years before Weinberg, in a popular-science book, The Fabric of the Cosmos [9], Brian Greene had also given a von Neumann-kind description of measurement:

> Stage one — the evolution of wavefunctions according to Schrödinger's equation — is mathematically rigorous, totally unambiguous, and fully accepted by the physics community. Stage two — the collapse of a wavefunction upon measurement — is, to the contrary, something that during the last eight decades has, at best, kept physics mildly bemused, and at worst, posed problems, puzzles and potential paradoxes that have devoured careers. The difficulty [...] is that according to Schrödinger's equation wave functions do *not* collapse. Wavefunction collapse is an add-on. It was introduced after Schrödinger discovered his equation, in an attempt to account for what experimenters actually see. [[9], p. 201.]

Thus, like Weinberg, Greene seems to be unconcerned by Bell's criticism that something could be wrong with the underlying dynamics. The Schrödinger equation that Greene refers to may be the dynamical equation of an outdated theory.

Like Weinberg, Greene was not satisfied and he gave very good arguments for his dissatisfaction:

[*Each*] *of the potential outcomes embodied in the wavefunction still vies for realization*. And so we are still wondering how one outcome "wins" and where the many other possibilities "go" when that actually happens. When a coin is tossed, [...] you can, in principle, *predict* whether it will land heads or tails. On closer inspection, then, precisely one outcome is determined by the details you initially overlooked. The same cannot be said in quantum physics. [...]

Much in the spirit of Bohr, some physicists believe that searching for such an explanation of how a single, definite outcome arises is misguided. These physicists argue that quantum mechanics, with its updating to include decoherence, is a sharply formulated theory whose predictions account for the behavior of laboratory measuring devices. And according to this view, *that* is the goal of science. To seek an explanation of *what's really going on*, to strive for an understanding of *how a particular outcome came to be*, to hunt for a level of *reality beyond detector readings and computer printouts* betrays an unreasonable intellectual greediness.

Many others, including me, have a different perspective. Explaining data *is* what science is about. But many physicists believe that science is also about embracing the theories data confirms and going further by using them to get maximal insight into the nature of reality. [[9], pp 212-213.]

This is also the view underlying this paper. In his book [5], Andrew Whitaker defined 'realism' in a very similar way:

[...] *realism* — the view that the observations that we make in science are related to a *real world* existing independently of our observations. When we analyse our results and produce theories, which are often mathematically abstract and elaborate, we are not merely trying to discover *correlations* that may enable us to predict further experimental results; we are attempting to gain information about this ultimate reality. [[5], pp 165-166.]

In our view, a one-world understanding is needed and it should be applicable not only to an ensemble of processes, but also to what happens in a single measurement.

Within our long quotation of Greene, he described the common understanding of classical coin-tossing. He then added: "The same cannot be said in quantum mechanics." However, in our discussion, we have concluded that, except when $\mu$ is already in an eigenstate of $\sigma_z$, the initial state of $A$ that wins the competition is also decisive for the measurement result. The measurement problem now, is to understand how this happens.

## 4.3 Unromantic approaches in Bell's typology

John Bell's book Speakable and Unspeakable in Quantum Mechanics [2] deals with quantum measurement as a research program to be completed by physical research. The book illuminates the problem from many different angles.

Bell had a typology, listing six approaches to the quantum measurement problem, three of them 'romantic' and three 'unromantic'. The Many-Worlds Interpretation is one of the romantic approaches.

In the present discussion, we stay within conventional (but up-to-date) quantum mechanics. With this, we place ourselves immediately in Bell's unromantic 'pragmatic' approach. Another unromantic approach, very close to Bell's thinking, is the 'non-linear/stochastic' approach with quantum diffusion as the most clear example. Actually, Bell had quite strong hopes that quantum diffusion could solve the measurement problem.

## 5. A MODERNIZED DISCUSSION

### 5.1. The measurement postulate

In the pragmatic approach, the measurement problem is to understand measurement in a single world as an interaction between the measured system $\mu$ and the system $A$, part of the measurement apparatus. We shall assume $A$ to be sufficiently small, so that, during the interaction, the combined system $\mu \cup A$ can be viewed as a closed quantum system.

First we shall briefly describe the elements of non-destructive measurement as presented by P.A.M. Dirac [10] (in particular Chapter II, p. 38). Such a measurement of $\sigma_z$ on $\mu$, with $\mu$ in the initial state $|\psi\rangle_\mu = \psi_+ |+\rangle_\mu + \psi_- |-\rangle_\mu$ ($|\psi_+|^2 + |\psi_-|^2 = 1$), (assuming $\psi_\pm \neq 0$) gives as result one of the eigenvalues +1 or −1, and the process takes $\mu$ into the corresponding eigenstate: when the result is +1, into the state $|+\rangle_\mu$, when the result is −1, into the state $|-\rangle_\mu$, with the probabilities $|\psi_+|^2$ and $|\psi_-|^2$ respectively.

As we have seen, the + result must obtain for the initial state of $A$ being in a certain group of initial states, let us call it $\Omega_+^{(i)}$; similarly, the − result obtains for an initial state of $A$ in another group of initial states $\Omega_-^{(i)}$. Thus $\Omega_+^{(i)}$ and $\Omega_-^{(i)}$ are subensembles of the whole ensemble of initial states $\Omega_A^{(i)}$ for $A$. Initial states belonging to neither of $\Omega_+^{(i)}$ or $\Omega_{-}^{(i)}$ must be so inefficient in leading to a final state that they do not contribute.

Denoting, for the moment, an initial state of $A$ by $|i,+\rangle_A$ in the + case and by $|i,-\rangle_A$ in the − case, we then have the situation that the initial states,

$$|\psi\rangle_\mu \otimes |i,+\rangle_A \quad \text{or} \quad |\psi\rangle_\mu \otimes |i,-\rangle_A, \tag{5.1}$$

lead to final states

$$|+\rangle_\mu \otimes |f,+\rangle_A \quad \text{or} \quad |-\rangle_\mu \otimes |f,-\rangle_A, \tag{5.2}$$

where it is understood that the final state of $A$, $|f,+\rangle_A$ or $|f,-\rangle_A$ is beginning to register a result. Moreover, the probabilities for the + and − events are $|\psi_+|^2$ and $|\psi_-|^2$, respectively.

The measurement problem is to find out <u>how</u> this can be so.

## 5.2. <u>Using scattering theory to describe $\mu A$-interaction</u>

We have argued that quantum mechanics should be understood as relativistic quantum mechanics, i.e., Quantum Field Theory. Different interaction strengths can be handled by allowing differences in transition amplitudes.

In this theory, the initial state of $A$ belongs to the ensemble $\Omega_A^{(i)}$; the states of this ensemble compete with the influence that they have over the transition rate.

We repeat part of the above quotation from Whitaker on von Neumann and measurement:

Von Neumann himself showed that a process of type 2 is *thermodynamically reversible*; it remains possible at the end of the process to restore the system to its initial state. For a process of type 1, no such possibility exists, and the process is called *thermodynamically irreversible*.

Actually, it is possible, and even necessary, to describe $\mu \cup A$ as a closed system with deterministic (and reversible) dynamics. <u>The statistical feature in quantum measurement comes from the statistics of the ensemble of possible initial states $\Omega_A^{(i)}$ of $A$ and their influence on the $\mu A$ transition rates.</u> After the $\mu A$-interaction, $A$ must also interact irreversibly with the main part of the measurement apparatus to make the result readable. Thus the entire type-1 process is irreversible but the first (and decisive) part of it, the $\mu A$-interaction, is reversible. Since the initial state of $A$ can <u>not</u> be known, the ensemble $\Omega_A^{(i)}$ of available initial states of $A$ and their influence on the final state, and hence on the result, must be handled statistically. Thus the statistical feature of measurement is <u>not an intrinsic property of quantum dynamics</u> ('God playing dice'); it has a <u>statistical-mechanics origin</u>. This source of statistics is of a kind that should be acceptable even according to the criteria that were used for physical theories by Albert Einstein.

The essential nature in the measurement process of the $\mu A$-interaction was emphasized by John Bell [2] who refers back to Niels Bohr on this point:

This word [measurement] very strongly suggests the ascertaining of some preexisting property of some thing, any instrument involved playing a purely passive role. Quantum experiments are just not like that, as we learned especially from Bohr. The result has to be regarded as a joint product of 'system' and 'apparatus', the complete experimental set-up. [[2] Chapter 17, p. 166]

Also in Chapter 23 of [2] 'Against measurement' with comments on earlier theoretical approaches, Bell emphasizes the importance of $\mu A$-interaction. The quantum system $S'$ in his terminology is our system $\mu \cup A$ and $R'$ is its environment. Bell argues that the reduction from the 'and' of the superposition to the 'or' of the possible results takes place <u>within</u> $S'$. He compares with earlier theories and writes:

The change is from a theory which speaks *only* of the results of external interventions on the quantum system, $S'$ in this discussion, to one in which that system is attributed *intrinsic properties*[.] [[2], p. 226.]

He adds that

[t]he point is strategically well chosen in that the predictions for results of 'measurements' across $S'/R'$ will still be the same. [[2], p. 226.]

Since the system $\mu$ is already known to be initially in the state $|\psi\rangle_\mu$, 'intrinsic properties' of $S'$ must refer to the initial state of the system $A$. This is very close to our discussion above with the Equations (5.1) and (5.2).

We note that in contrast to this, in each term of the superposition resulting in RSF/MWI, the apparatus has a purely passive role.

We shall now use the book Theory of Photons and Electrons by J.M. Jauch and F. Rohrlich [11], published in 1955, about two years before the publication of Everett's paper. We have chosen to use as our starting point Equation (8-40) of the book. The transition rate from an initial state $|i\rangle$ to a final

state $|f\rangle$ is

$$(2\pi)^{-1} S_f \bar{S}_i \delta(P_f - P_i)|\langle f|M|i\rangle|^2, \qquad (5.3)$$

where $\langle f|M|i\rangle$ is defined in the same book, 2 pages earlier, from the scattering operator $S$ through

$$\langle f|S-1|i\rangle = \delta(P_f - P_i)\langle f|M|i\rangle. \qquad (5.4)$$

It should be pointed out that $M|i\rangle$ is not a normalized state. In Equation JR (8-40), $S_f$ and $\bar{S}_i$ denote a summation over observationally indistinguishable final states $|f\rangle$, and a similar averaging over initial states $|i\rangle$, respectively.

The averaging $\bar{S}_i$ reminds us that $A$ has a large ensemble $\Omega_A^{(i)}$ of available initial states in which two subensembles, $\Omega_+^{(i)}$ and $\Omega_-^{(i)}$, have very special roles. The states in these subensembles have the highest transition rates and therefore they dominate the transitions to final states.

In principle, Everett and Wheeler had the possibility to use JR (8-40) rather than the old Schrödinger dynamics of Everett's paper. They did not.

It can be useful to have an expression for the density matrix of the final state. We first note that $|i\rangle$ is a state different from $|f\rangle$, because of the impact made by $\mu$ on $A$. Therefore $\langle f|i\rangle = 0$ and $\langle f|S|i\rangle = \langle f|S-1|i\rangle$, and JR (8-40) is adequate to describe the whole process. For given 4-momentum $P_i = P_f$, we then have

$$\rho_{fg} = \frac{\langle f|M|i\rangle\langle i|M^\dagger|g\rangle}{\sum_{f'}|\langle f'|M|i\rangle|^2} = \frac{\langle f|M|i\rangle\langle i|M^\dagger|g\rangle}{\sum_{f'}\langle f'|M|i\rangle\langle i|M^\dagger|f'\rangle} \quad (\text{Tr}\rho = 1). \qquad (5.5)$$

We see that the final-state density matrix is non-linear in the density matrix for the initial state $|i\rangle\langle i|$; this is necessary for normalization.

One can show that, dynamically, this normalization is closely related to the reversibility of Quantum Field Theory. (In an alternative description, $\mu$ is created in its initial state $|\psi\rangle_\mu$ by an external source and absorbed after the $\mu A$-interaction in an eigenstate of $\sigma_z$, $|+\rangle_\mu$ or $|-\rangle_\mu$, by an external sink. Everything can be described explicitly in a Feynman-diagram representation and summed to all orders of perturbation theory. In this alternative description, both unitarity of the S-matrix and reversibility are explicitly manifest. This has been worked out in detail in a previous article [12].)

### 5.3. Initial and final states of the $\mu A$-scattering process

Initial states for $\mu$, $|+\rangle_\mu$, $|-\rangle_\mu$ and $|\psi\rangle_\mu$, have already been introduced. To denote general initial states of $A$, we use only variables that are of importance for the transition to a final state,

$$|\underline{\varepsilon};0\rangle_A, \quad \underline{\varepsilon} = (\varepsilon_1, \varepsilon_2, ..., \varepsilon_N); \quad \langle\underline{\varepsilon};0|\underline{\varepsilon}';0\rangle = \delta_{\underline{\varepsilon}\underline{\varepsilon}'}. \tag{5.6}$$

We now use a more flexible notation (than $|i,+\rangle_A$ and $|i,-\rangle_A$ in Section 5.1 above) for the initial states of $A$ with '0' denoting readiness of $A$ to receive an impact from $\mu$. The variables $\underline{\varepsilon}$ are standardized discrete variables to be defined later. The states $|\underline{\varepsilon};0\rangle_A$ make up the ensemble $\Omega_A^{(i)}$. The initial state to be considered is thus

$$|i\rangle = |\psi\rangle_\mu \otimes |\underline{\varepsilon};0\rangle_A, \tag{5.7}$$

with $|\psi\rangle_\mu$ given as the superposition (3.1).

The scattering process takes the <u>initial states</u>

$$|+\rangle_\mu \otimes |\underline{\varepsilon};0\rangle_A \quad \text{and} \quad |-\rangle_\mu \otimes |\underline{\varepsilon};0\rangle_A \tag{5.8}$$

into the <u>non-normalized final states</u>

$$M|+\rangle_\mu \otimes |\underline{\varepsilon};0\rangle_A = b^{(+)}(\underline{\varepsilon})|+\rangle_\mu \otimes |\beta_+(\underline{\varepsilon});+\rangle_A \tag{5.9}$$

and

$$M|-\rangle_\mu \otimes |\underline{\varepsilon};0\rangle_A = b^{(-)}(\underline{\varepsilon})|-\rangle_\mu \otimes |\beta_-(\underline{\varepsilon});-\rangle_A, \tag{5.10}$$

where $b^{(+)}(\underline{\varepsilon})$ and $b^{(-)}(\underline{\varepsilon})$ are transition amplitudes which can depend strongly on the initial states $|\underline{\varepsilon};0\rangle_A$ of $A$. In the final states, '+' and '−' replacing '0' denote beginning registration of a result.

When the initial state is the superposition (5.7), the whole interaction process leading to a non-normalized final state

is (see Figure 1)

$$|\psi\rangle_\mu \otimes |\varepsilon;0\rangle_A \to M|\psi\rangle_\mu \otimes |\varepsilon;0\rangle_A =$$
$$= \psi_+ b^{(+)}(\varepsilon)|+\rangle_\mu \otimes |\beta_+(\varepsilon);+\rangle_A + \psi_- b^{(-)}(\varepsilon)|-\rangle_\mu \otimes |\beta_-(\varepsilon);-\rangle_A. \quad (5.11)$$

The <u>big difference</u> between this and the corresponding equations of the 1930s (see Equation (2.5)) that led to von Neumann's dilemma is the appearance of the <u>transition amplitudes</u> $b^{(+)}(\varepsilon)$ and $b^{(-)}(\varepsilon)$, describing the strengths of the different processes. This is crucial; the final state is <u>not</u> normalized and we can expect its + and − components to vary widely with <u>ε</u>.

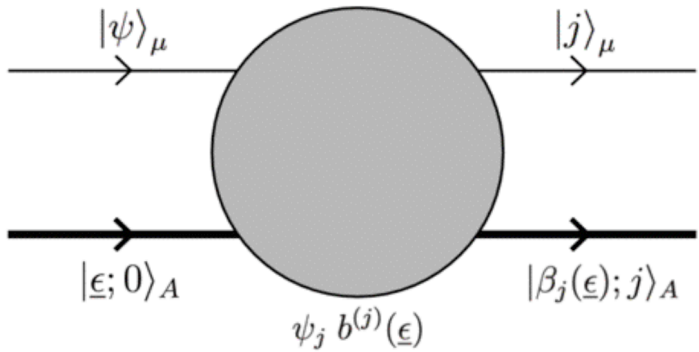

**Figure 1.** Schematic Feynman diagram for a transition from the initial state $|\psi\rangle_\mu \otimes |\varepsilon,0\rangle_A$ to the final state $|j\rangle_\mu \otimes |\beta_j(\varepsilon),j\rangle_A$, $j = \pm$. The transition amplitude $\psi_j b^{(j)}(\varepsilon)$ depends both on $\psi_j$ and on microscopic details of the initial state $|\varepsilon, 0\rangle_A$ of the system $A$.

For the final state reached, to be a representative final state, it should look like (5.1) or (5.2), i.e., we must have <u>either</u> that

$$|b^{(+)}(\varepsilon)|^2 \gg |b^{(-)}(\varepsilon)|^2, \quad (5.12+)$$

a state in the subensemble $\Omega_+^{(i)}$, giving a + result, <u>or</u> that

$$|b^{(-)}(\varepsilon)|^2 \gg |b^{(+)}(\varepsilon)|^2, \quad (5.12-)$$

a state in the subensemble $\Omega_-^{(i)}$, giving a − result. All other member states of the ensemble $\Omega_A^{(i)}$, corresponding to values of <u>ε</u> not fulfilling any of the conditions (5.12+) or (5.12−), must somehow lead to final states with a negligible transition rate

$(2\pi)^{-1}w$ where

$$w = |\psi_+|^2 |b^{(+)}(\underline{\varepsilon})|^2 + |\psi_-|^2 |b^{(-)}(\underline{\varepsilon})|^2. \tag{5.13}$$

If we <u>find the mechanism providing such a selective situation</u>, then von Neumann's dilemma disappears completely. Reduction of the measured system to an eigenstate of the observable is no longer impossible. The system $\mu$ can come arbitrarily close to an eigenstate of $\sigma_z$, and we may have an explanation for the bifurcation taking place in measurement, resulting in + <u>or</u> —.

## 5.4. The non-linear/stochastic approach

The first development in the direction of a quantum-mechanical analysis of measurement came with Quantum Diffusion [13] in the 1980s and early 1990s. This theory looks very non-linear. However, quantum diffusion can be shown to be simply an alternative perspective of $\mu A$-scattering, focusing on how the state of $\mu$ is influenced by a stepwise extended system $A$. It can be derived from our model within the pragmatic approach to be presented in the next subsection. Thus despite its non-linearity, quantum diffusion is a consequence of linear quantum theory [14]. We can therefore consider the non-linear/stochastic approach to be an alternative version of the pragmatic approach.

## 5.5. Our model within the 'pragmatic' approach

Bell described the pragmatic approach in this way [2]:

The pragmatic philosophy is, I think, consciously or unconsciously the working philosophy of all who work with quantum theory in a practical way... when so working. [[2], p. 189.]

and the non-linear/stochastic approach as follows:

This possible way ahead is unromantic in that it requires mathematical work by theoretical physicists, rather than interpretation by philosophers, and does not promise lessons in philosophy for philosophers. [[2], p. 190.]

Our model of measurement within the pragmatic framework has been presented elsewhere [12]. The crucial features of the initial states of $A$, are their influence on the transition amplitudes. We analyze statistically the initial states of $A$ and how some of them group themselves in $\Omega_+^{(i)}$ and $\Omega_-^{(i)}$, according to their influence on the transition rates.

We construct our model as follows. We successively increase the number $N$ of variables in $\underline{\varepsilon}$ in a thought procedure of stepwise extending the system $A$ and consider how this influences the amplitudes $b^{(+)}(\underline{\varepsilon})$ and $b^{(-)}(\underline{\varepsilon})$. Only changes that influence the two amplitudes differently, need to be considered. In the mean there should be no difference between the changes for + and −. For simplicity, we have chosen to change the amplitudes for each step by standard factors with modulus close to unity in such a way that the means of the squared moduli of the amplitudes do not change. When $N$ is increased from $n-1$ to $n$, the variable $\varepsilon_n$ enters, and we let $b^{(j)}(\underline{\varepsilon})$ change by a factor

$$\left(1 + \tfrac{1}{2} j \varepsilon_n \kappa - \tfrac{1}{8} \kappa^2\right) e^{\tfrac{1}{2} i j \phi_n} \quad (j = \pm), \tag{5.14}$$

where $\phi_n$ is a phase and

$$\begin{aligned} &0 < \kappa << 1; \\ &\varepsilon_n = \pm 1;\ \langle\langle \varepsilon_n \rangle\rangle = 0,\ \langle\langle \varepsilon_n \varepsilon_{n'} \rangle\rangle = \delta_{nn'}. \end{aligned} \tag{5.15}$$

Here $\kappa$ determines the size of a standard step. For $\varepsilon_n = +1$, the + result is strengthened and the − result is weakened; for $\varepsilon_n = -1$, the opposite happens. The mean $\langle\langle\ \rangle\rangle$ is the mean over the ensemble $\Omega_A^{(i)}$ of available initial states of $A$.

Our construction of $b^{(j)}(\underline{\varepsilon})$ is probably the simplest possible, given the conditions that the steps are small and independent and that they are free from systematic bias in favour of one of the results. This construction is <u>the crucial assumption of our model</u>.

In our setting up of $A$, we are thus using standard steps. This means that we can expect the net number of steps in the + direction to have special significance as an aggregate measure. We thus define

$$Z(\underline{\varepsilon}) = \sum_{n=1}^{N} \varepsilon_n,\ Z = -N,\ -N+2,\ ...,\ N-2,\ N,\ \langle\langle Z \rangle\rangle = 0,\ \langle\langle Z^2 \rangle\rangle = N. \tag{5.16}$$

A positive value of $Z$ expresses a relative advantage given to the + result and a negative value expresses a relative advantage given to the − result.

Explicitly, we get the squared moduli of ± amplitudes,

$$\left|b^{(j)}(\underline{\varepsilon})\right|^2 = \prod_{n=1}^{N}(1 + j\varepsilon_n \kappa) = e^{jZ\kappa - \tfrac{1}{2}\Xi},\ \left\langle\left\langle \left|b^{(j)}(\underline{\varepsilon})\right|^2 \right\rangle\right\rangle = 1;\ (j = \pm)., \tag{5.17}$$

where

$$\Xi = N\kappa^2 \tag{5.18}$$

is a measure of the total variance, independent of $\underline{\varepsilon}$. We shall be interested in the limit of large $\Xi$.

It will be convenient to use instead of $Z$ defined in (5.16), the following, more suitably normalized, aggregate variable for the initial state of $A$,

$$Y = Y(\underline{\varepsilon}) = \frac{\kappa}{\Xi} Z(\underline{\varepsilon}) = \frac{Z(\underline{\varepsilon})}{N\kappa} \quad \left( \langle\langle Y \rangle\rangle = 0, \quad \langle\langle Y^2 \rangle\rangle = \frac{1}{\Xi} \right). \tag{5.19}$$

$Y$ is a discrete variable, but with increasing $\Xi$, neighbouring values of $Y$ become very close, and it is therefore convenient to think of $Y$ as a continuous variable. It will be natural to consider $Y(\underline{\varepsilon})$ as applicable not only to the initial state $|\underline{\varepsilon};0\rangle_A$ of $A$, but also to the corresponding final states, the daughter states $|\beta_+(\underline{\varepsilon});+\rangle_A$ and $|\beta_-(\underline{\varepsilon});-\rangle_A$.

Clearly, each $\varepsilon_n$ is a complicated function of positions, momenta and internal degrees of freedom of those particles that become included in the $n$th step of the extension of $A$. We cannot and do not have to deal with these complications. What matters to us is the step size $\kappa$ and the lack of bias, i.e., the total ambivalence of $\varepsilon_n$ in each step, whether to favour + or -.

The transition amplitudes as functions of $Y$ and $\Phi = \sum_{n=1}^{N} \phi_n$ are

$$b^{(j)}(Y,\Phi) = \prod_{n=1}^{N} \left(1 + \tfrac{1}{2} j\varepsilon_n \kappa - \tfrac{1}{8}\kappa^2 \right) e^{\tfrac{1}{2}ij\phi_n} = e^{\tfrac{1}{2}\Xi(jY - \tfrac{1}{2}) + \tfrac{1}{2}ij\Phi};$$
$$\langle\langle |b^{(j)}|^2 \rangle\rangle = 1 \quad (j = \pm). \tag{5.20}$$

The total transition rate (for an initial state $|i\rangle = |\psi\rangle_\mu \otimes |\underline{\varepsilon};0\rangle_A$ with $Y(\underline{\varepsilon}) = Y$, is $(2\pi)^{-1} w(Y)$ with

$$w(Y) = |\psi_+|^2 |b^{(+)}|^2 + |\psi_-|^2 |b^{(-)}|^2 = |\psi_+|^2 e^{\Xi(Y - \tfrac{1}{2})} + |\psi_-|^2 e^{\Xi(-Y - \tfrac{1}{2})}. \tag{5.21}$$

The mean of $w(Y)$ among initial states is

$$\langle\langle w(Y) \rangle\rangle = 1. \tag{5.22}$$

The transition rate to a final state with $\mu$ in the state $|j\rangle_\mu$ ($j = \pm$) is the corresponding term of (5.21),

$$(2\pi)^{-1}|\psi_j|^2 e^{\Xi(jY-\frac{1}{2})}. \tag{5.23}$$

For an initial state $|\varepsilon;0\rangle_A$ with $Y(\varepsilon) = Y$ given, the <u>final-state density matrix</u> in the basis $|j\rangle_\mu \otimes |\beta_j(\varepsilon);j\rangle_A$ ($j = \pm$), written as a function of $Y$ and the phase $\Phi$, is

$$\rho(Y,\Phi) = \frac{1}{|\psi_+|^2 e^{\Xi Y} + |\psi_-|^2 e^{-\Xi Y}} \begin{pmatrix} |\psi_+|^2 e^{\Xi Y} & \psi_+ \psi_-^* e^{i\Phi} \\ \psi_- \psi_+^* e^{-i\Phi} & |\psi_-|^2 e^{-\Xi Y} \end{pmatrix}. \tag{5.24}$$

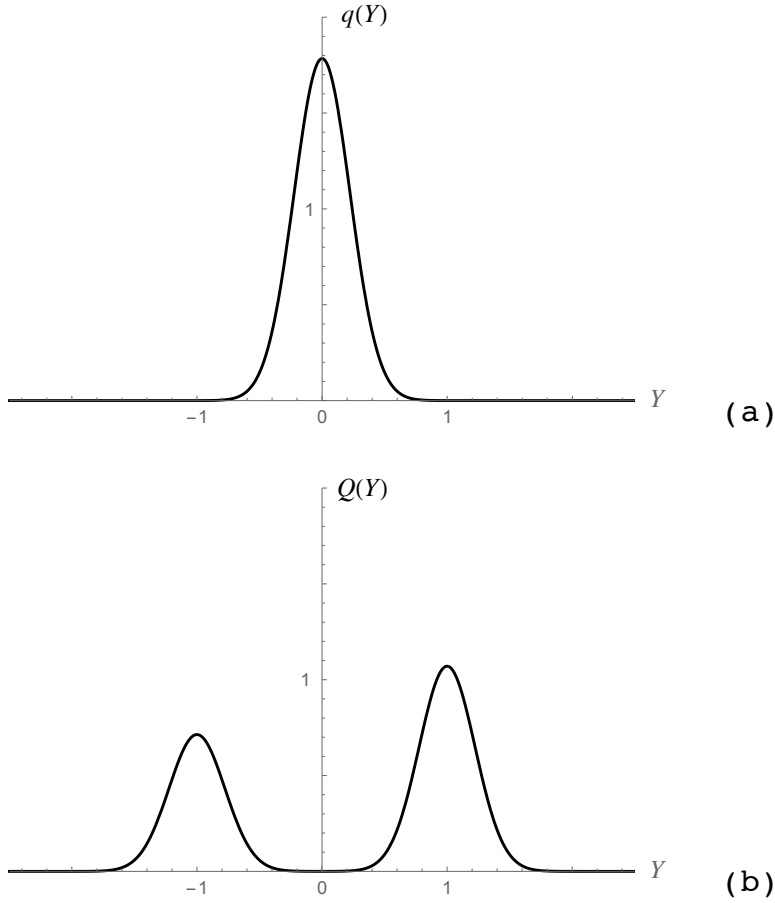

**Figure 2.** The distribution $q(Y)$ over $Y$ for the initial states of the system $A$ (**Figure 2a**), and the corresponding distribution $Q(Y)$ for the final states of $\mu \cup A$, Eq. (5.26), centered around $Y = 1$ and $Y = -1$ (**Figure 2b**), based on the values $\Xi = 20$, and $(|\psi_+|^2, |\psi_-|^2) = (0.6, 0.4)$.

We turn to the distribution over $Y$ for the states of $A$, Among the initial states of $\Omega_A^{(i)}$, the mean and variance of $Y$ are given by (5.19). The distribution is narrow in the limit of large $\Xi$ and well represented by the Gaussian (Figure 2a)

$$q(Y) = \sqrt{\frac{\Xi}{2\pi}} e^{-\frac{1}{2}\Xi Y^2} . \tag{5.25}$$

For the final states, the strong dependence on $Y$ of the transition rates changes the picture completely. To get the distribution over the final states, the initial-state Gaussian $q(Y)$ has to be multiplied by the normalized, transition-rate factor (5.21). The resulting <u>final-state distribution</u> over $Y$ or, what is the same, the <u>distribution of transition processes</u> over $Y$, is

$$Q(Y) = q(Y)w(Y) = \sqrt{\frac{\Xi}{2\pi}} e^{-\frac{1}{2}\Xi Y^2} \left( |\psi_+|^2 e^{\Xi(Y-\frac{1}{2})} + |\psi_-|^2 e^{\Xi(-Y-\frac{1}{2})} \right) =$$
$$= |\psi_+|^2 q(Y-1) + |\psi_-|^2 q(Y+1), \tag{5.26}$$

i.e., the sum of two Gaussians, one for + and one for −, weighed by $|\psi_+|^2$ and $|\psi_-|^2$, respectively. The + term is centered around $Y=1$ and the − term is centered around $Y=-1$. The transition rates depend strongly on the initial states. Thus the transition process involves a <u>selection of states from two, well separated, peripheral subensembles of</u> $\Omega_A^{(i)}$, $\Omega_+^{(i)}$ and $\Omega_-^{(i)}$, with internal distributions $q(Y-1)$ and $q(Y+1)$, respectively (Figure 2b). The coefficients $|\psi_+|^2$ and $|\psi_-|^2$ confirm the <u>Born rule</u>.

In the limit of large $\Xi$, the density matrix for the final state (5.24) is

$$\rho(1,\Phi) = \begin{pmatrix} 1 & 0 \\ 0 & 0 \end{pmatrix} \text{ or } \rho(-1,\Phi) = \begin{pmatrix} 0 & 0 \\ 0 & 1 \end{pmatrix}, \tag{5.27}$$

as expected. The mean of the final-state density matrix for $\mu$ is

$$\int dY Q(Y)\rho(Y,\Phi) = \begin{pmatrix} |\psi_+|^2 & 0 \\ 0 & |\psi_-|^2 \end{pmatrix}, \tag{5.28}$$

as also expected.

## 6. CONCLUDING REMARKS

### 6.1. Romantic vs unromantic approaches

To discuss work within the pragmatic view on measurement with physics colleagues is very difficult due to a general lack of interest, based on the widespread belief that quantum mechanics can be used for analysis a measurement only in a MWI, <u>not</u> in a one-world context.

John Bell wrote a comment on the advantage of the 'romantic worlds' in gaining a wider attention:

It is easy to understand the attraction of the three romantic worlds for journalists, trying to hold the attention of the man in the street. The opposite of truth is also a truth! Scientists say that matter is not possible without mind! All possible worlds are actual worlds! Wow! And the journalists can write these things with good consciences, for things like this have been said ... out of working hours ... by great physicists. For my part, I never got the hang of complementarity, and remain unhappy about contradictions. As regards mind, I am fully convinced that it has a central place in the ultimate nature of reality. But I am very doubtful that contemporary physics has reached so deeply down that that idea will soon be professionally fruitful. For our generation I think we can more profitably seek Bohr's necessary 'classical terms' in ordinary macroscopic objects, rather than in the mind of the observer. The 'many world interpretation' seems to me an extravagant, and above all an extravagantly vague, hypothesis. [[2] Chapter 20, pp. 193-194.]

To let one's view on human activities determine one's understanding of the natural world is what characterizes a <u>romantic</u> view. In quantum mechanics, the epistemological problems concerning measurement have tempted physicists to postulate statements about the nature of reality, as Bell described in his parodic characterization above. Ontological statements are made that are not founded on experience but in a feeling of what is needed to circumvent the epistemological problems.

As we have repeatedly indicated in quotations and comments, Bell argued for another way, based on a realist ambition. He showed that the old dynamics used by RSF and MWI leads to a situation which is not characteristic for quantum mechanics because of its lack of reversibility.

Referring to Bohr, Bell stressed that the result of a measurement is a joint product of the interacting systems in $S'$, $\mu$ and $A$. Moreover he stated his expectation that the bifurcation of measurement, the splitting of the 'and' of the superposition into the 'or' of the different outcomes, is a result of the intrinsic properties of $S'$, i.e., of $\mu$ and $A$. Our work, as we have reviewed it here, follows these ideas from Bell.

The theories involved are quantum mechanics and statistical mechanics, applied to unknown initial states of the system $A$, most of them inactive but some of them very active. The statistical features of quantum measurement have their origin in this statistics. This is far from an inherently capricious nature or an interfering human mind, also far from allowing everything to happen in separate parallel worlds.

We have kept our theoretical discussion of measurement at a very general level. For a clear understanding, it will be necessary to combine this with detailed analysis of concrete measurement situations. Already in the early days of quantum mechanics, Mott [15] started this kind of analysis; it is good to see that this kind of work continues [16].

## 6.2. The difficulty of the physics community to see the generality of relativistic quantum mechanics

The relativistic quantum mechanics needed for our work on measurement that has been reviewed here, was available already at the time of Everett's paper. Both RSF and MWI were based on a version of quantum theory that was already obsolete at that time. As we have shown, scattering theory of Quantum Field Theory could have opened for a more conventional and more fruitful analysis of quantum measurement. The tradition initiated with RSF and MWI was unnecessary already from the beginning.

This gives me no reason to blame Everett [3] or DeWitt [4] or the editors of Reviews of Modern Physics or Physics Today, nor any person following in the RSF/MWI tradition. The failure to see the new possibilities was a failure of the entire physics community, a failure within the common culture of physics. In the Introduction, we quoted Feynman, who could not see the great relevance of his own work to solving the measurement problem.

I was myself interested in the measurement problem all since my first studies of quantum mechanics in 1956. I remember reading Everett's paper in the peaceful CERN Library, probably in 1960. At that time, in my daily work, I was using the book by Jauch and Rohrlich [11] as a handbook for calculations in Quantum Electrodynamics. I was also caught in the culture of the time; I did not see any connection between measurement and the theory that I was using.

John Bell joined the theoretical staff of CERN about this time. He told me about his ambition to analyze quantum measurement.

No theory in science has a deeper or wider explanatory power than quantum mechanics. But still quantum mechanics is described as weird, spooky, mysterious, unintelligible. In these cases one rather speaks about the old proto-quantum mechanics of the 1930s. The quantum mechanics proper has, in many cases not been carefully examined.

This ambiguous attitude towards quantum mechanics should be abandoned. Quantum mechanics should be examined and judged in its developed form. The step from non-relativistic quantum mechanics to relativistic quantum mechanics should finally be taken. We can learn from the situation in the life sciences in the first decade of the 20th century. At this time, it became necessary to take the decisive step from vitalism and recognize that metabolic processes could be understood as chemical reactions. 110 years after biochemistry, physics should be ready take the corresponding step.

Quantum mechanics should be a general framework for physical analysis. There may be a lot to be discovered. Now is the time to give Quantum Mechanics a fair chance!

Acknowledgements

I thank Andrew Whitaker and Kristian Lindgren for discussions on physical issues and Magdalena Eriksson and Sven-Eric Liedman for discussions on the historical aspects.

ADDENDUM:
THE VON NEUMANN TRADITION AND BELL'S PRAGMATIC APPROACH

There is a common understanding of what quantum mechanics means concerning physical systems and processes. This is vital for the ongoing development in physics. There is however a lack of understanding of how quantum measurement functions. This causes an uncertainty also about the exact meaning of the basic concepts of the theory. Every student of quantum mechanics must make a personal picture of these basic concepts and their relations.

Therefore, even if there is agreement about most things in quantum mechanics and good traditions in communication and recording, there can be big differences in the understanding at a conceptual level. A physicist's picture of reality therefore contains, beside a common component, a component of personal choice. The physicist has the option to accept one out of a few standard theories, to form a strictly personal view, or to leave the question open. This situation which makes the view of basic concepts to some extent have the character of belief, can make the discussion on conceptual matters in quantum mechanics quite difficult.

One can follow Niels Bohr and consider quantum mechanics as a theory that is inherently statistical. This is one way to solve the measurement problem by rejecting it. However to introduce statistics at a fundamental level in science is a great step, and Einstein was strongly against it. The MWI (See Section 2.2) is another way to reject the problem and instead introduce an 'extravagant' (Bell's word) cosmology.

The descriptions of the measurement problem, given by Steven Weinberg and Brian Greene (see Section 4 above), follow largely von Neumann's dilemma, Schrödinger's cat idea, and the problem description of RSF/MWI. We shall name this line of thought the "von Neumann tradition". It is probably correct to consider this as mainstream among physicists at the present time.

A totally different attitude (and in the long perspective more traditional) is that of John Bell in his book [2]: to view the measurement problem as a research problem to be solved within physics, like other problems, and to look for theoretical ideas and experimental information that can be relevant for this problem. His criticism of Everett-DeWitt as working with a theory that does not look like quantum mechanics — "branching but not debranching" — was given in this spirit. Bell suggested that quantum theory instead could be represented by Feynman's "sum over paths".

Bell also said that we must consider the measurement result as an effect of $\mu A$-interaction, not just as an effect of $\mu$ acting on a passive system *A* (see Section 5.2). This is also a clear deviation from the von Neumann tradition. Moreover, one of Bell's unromantic approaches is the "pragmatic approach", to solve the problem through continued regular work in quantum mechanics. We shall use the name for the whole of this line of action: "Bell's pragmatic approach".

Close to Feynman's sum over paths is relativistic quantum mechanics, i.e., Quantum Field Theory, where transition amplitudes and transition rates become key concepts. We have chosen to use relativistic quantum mechanics as representing quantum mechanics. I have asked Andrew Whitaker who knows the history of the measurement problem [5], if he knows of any other attempt to consider relativistic quantum mechanics as the quantum mechanics to be used in connection with the measurement problem. His answer was: No.

If I may be allowed to view the first few steps of our work as a continuation of the pragmatic approach (see Sections 4.3 and 5.3), then the pragmatic approach immediately goes against the von Neumann tradition.

In the next section, I shall discuss the von Neumann tradition and the pragmatic approach in relation to each other. For this, I shall use definitions from Alan Sokal's [17] discussion about science and pseudoscience.

A1. Mainstream science and pseudoscience?

Alan Sokal [17] has made studies on pseudoscience and on postmodernism (extreme belief and extreme skepticism) and he has found that these two can have common interests and sometimes support each other in their disagreement with realism.

Of main interest here are his definitions of pseudoscience and of postmodernism. We start here with pseudoscience.

Sokal used the following definition:

Pseudoscience is any body of thought (along with its associated justifications and advocates) that

(a) makes assertions about real or alleged phenomena and/or real or alleged causal relations that mainstream science justifiably considers to be utterly implausible, and

(b) attempts to support these assertions through types of
    argumentation or evidence that fall far short of the
    logical and evidentiary standards of mainstream science.

The crucial terms in this definition are 'mainstream science'
and 'justifiably'. The personal component of the understanding
of quantum mechanics can play a role here. Beyond the personal
level, it can lead to common convictions about the right way to
view the conceptual basis of quantum mechanics. There is a risk
that arbitrary dogmas are formed and get recognized as part of
mainstream science, for instance the idea that quantum
mechanics cannot be used to analyze measurement.

The von Neumann tradition is based on the non-relativistic
quantum mechanics of the 1930s. In actual physics, it has now
been replaced by relativistic quantum mechanics; no response
seems to have been given from proponents of the von Neumann
tradition to Bell's criticism. Largely, it seems that the von
Neumann tradition has conquered the role of mainstream physics
but to me it looks like a set of arbitrary dogmas.

John Bell was very highly regarded within mainstream physics
and his criticism should be seriously considered. His openness
with respect to analyzing measurement is the common openness of
science. The basic theory of Bell's pragmatic approach is
simply quantum mechanics, i.e., the theory commonly used by
physicists working on quantum-mechanical problems. Our addition
is only to include standard concepts of the scattering theory
of Quantum Field Theory.

For these reasons, in a <u>long-term perspective</u>, Bell's pragmatic
approach can be expected to be more stably mainstream than the
von Neumann tradition, since it rests on a firmer basis. If so,
MWI is not only superfluous but can justifiably be
characterized as "utterly improbable".

This is my view. The final judgement I must leave to others.

A2. <u>The Many-Worlds Interpretation and postmodernism</u>

Sokal's definition of <u>postmodernism</u> is the following:

> an intellectual current characterized by the more-or-less
> explicit rejection of the rationalist tradition of the
> Enlightenment, by theoretical discourses disconnected from
> any empirical test, and by cognitive and cultural
> relativism that regards science as nothing more than a
> "narration", a "myth" or a social construction among many
> others.

Let us make a description of MWI:

MWI is a world view that regards our experiences of reality as nothing more than one narrative about our world among innumerably many other narratives in a world of worlds. A narrative exists only in one world and cannot be communicated to another world.

This is probably close to what also an MWI proponent could accept.

Then what do postmodernists see in MWI? They see MWI as a world view structured in a way that is similar to postmodernism and therefore useful to support it. Even if postmodernists do not value science, they can value the prestige of science, as shown by Sokal [18]. Therefore, MWI can be felt as a scientific support for postmodernism.

The cultural attraction of romantic views like MWI was described earlier in quotations of Weinberg and Bell.
- - -

The purpose of this addendum is to show that the prospect for Bell's pragmatic approach, is quite good. The 'deconstruction' of quantum mechanics in the von Neumann tradition is only a deconstruction of the non-relativistic theory of the 1930s. More physics colleagues should go into Bell's pragmatic approach. We would, of course, welcome a critical evaluation of our work [12]. I also think that new perspectives and new ideas are needed.

As I have indicated above (Section 5.4), and as we plan to show more in detail [14], quantum diffusion can be viewed as part of the pragmatic approach, as the quantum mechanics for open systems [13]. (Because of inevitable entanglement with the interacting system, this is not simple; it needs further analysis).

Anyhow, we physicists should finally free ourselves from the von Neumann tradition and abandon MWI and continue work within Bell's pragmatic approach.

Romantic approaches and mystification of quantum mechanics may lead to public interest, but in the long run, clarity and understanding, developed within physics, are much more valuable to the human society.